\documentclass[%
reprint,
superscriptaddress,
 amsmath,amssymb,
 aps,
 prd,
]{revtex4-1}

\usepackage{graphicx}
\usepackage[mathlines]{lineno}

\newcommand\vtick{\ensuremath{^\prime}}									
\newcommand*\ops{{\ensuremath{\text{o-Ps}}}}
\newcommand*\pps{{\ensuremath{\text{p-Ps}}}}
\newcommand*\opsM{{\ensuremath{\textnormal{o-Ps\vtick}}}}
\newcommand*\gammaM{{\ensuremath{\gamma\textnormal{\vtick}}}}

\usepackage[detect-all]{siunitx}
	\DeclareSIUnit\gauss{G}											
	\DeclareSIUnit\positron{\text{\ensuremath{\text{e}^+}}}			
	\DeclareSIUnit\electron{\text{\ensuremath{\text{e}^-}}}			
\usepackage{physics}

\usepackage[caption=false]{subfig}

\usepackage{multirow}

\usepackage{hyperref}
\hypersetup
{
    pdfauthor = {Carlos Vigo},
	pdftitle = {First search for invisible decays of ortho-positronium confined in a vacuum cavity},
	pdfkeywords = {Positronium, Mirror Matter},
}

\begin{document}


\title{First search for invisible decays of ortho-positronium confined in a vacuum cavity}

\author{C.~Vigo}
\author{L.~Gerchow}
\affiliation{Institute for Particle Physics and Astrophysics, ETH Zurich, 8093 Zurich, Switzerland}
\author{L.~Liszkay}
\affiliation{IRFU, CEA, University Paris-Saclay F-91191 Gif-sur-Yvette Cedex, France}
\author{A.~Rubbia}
\author{P.~Crivelli}
\email{paolo.crivelli@cern.ch}
\affiliation{Institute for Particle Physics and Astrophysics, ETH Zurich, 8093 Zurich, Switzerland}

\date{\today}

\begin{abstract}
The experimental setup and results of the first search for invisible decays of ortho-positronium (\ops) confined in a vacuum cavity are reported. No evidence of invisible decays at a level $\text{Br}\left(\ops\to\text{invisible}\right) < \num{5.9E-4}$ (\SI{90}{\percent} C.~L.) was found. This decay channel is predicted in Hidden Sector models such as the Mirror Matter (MM), which could be a candidate for Dark Matter. Analyzed within the MM context, this result provides an upper limit on the kinetic mixing strength between ordinary and mirror photons of $\varepsilon < \num{3.1E-7}$ (\SI{90}{\percent} C.~L.). This limit was obtained for the first time in vacuum free of systematic effects due to collisions with matter.
\end{abstract}

\pacs{Valid PACS appear here}
\maketitle

\section{\label{sec:Intro}Introduction}

The origin of Dark Matter is a question of great importance for both cosmology and particle physics. The existence of Dark Matter has very strong evidence from cosmological observations~\cite{Bertone2005} at many different scales, e.g.~rotational curves of galaxies~\cite{Begeman1991}, gravitational lensing~\cite{Massey2010} and the cosmic microwave background CMB spectrum. The latest Planck Mission results~\cite{Planck2015} provide an accurate estimate of the abundance of baryons $\Omega_\text{b} = \num{0.048}$ and cold matter $\Omega_\text{c} = \num{0.258}$, leading to an abundance of cold matter $\sim\num{5}$ times larger than ordinary matter. The explanation of such observations is one of the strongest hints of the existence of new physics.

Many Dark Matter candidates have been hypothesized so far, the most relevant being sterile neutrinos, axions and supersymmetric particles (see Ref.~\cite{PDG2016} for a detailed, recent review). Supersymmetry is theoretically very attractive since in addition of providing a good candidate for DM (the Lightest Supersymmetric Particle, LSP), it could potentially solve the hierarchy problem~\cite{Martin1998} and grant unification of gauge couplings at high energies~\cite{Ellis1992}, necessary for Grand Unified Theories (GUT). However, all experimental searches have failed to provide any evidence of supersymmetry so far~\cite{PDG2016,ATLAS_SUSY_2015,CMS_SUSY_2015}.

Another interesting approach is the concept of a Hidden Sector (HS) consisting of a $\text{SU}\left(\num{3}\right)_\text{C}\otimes\text{SU}\left(\num{2}\right)_\text{L}\otimes\text{U}\left(\num{1}\right)_\text{Y}$ singlet field~\cite{Holdom1986}. These models extend the SM by introducing a sector which transforms under the new gauge group. Among the many HS scenarios, the Mirror Sector is a particularly interesting one, featuring a natural Dark Matter candidate (actually a whole set of candidates) and feasible experimental signatures via oscillations of ordinary matter into the HS, as well as a possible explanation of the anomaly reported by the DAMA Collaboration~\cite{Bernabei2016}.

\subsection{\label{subsec:Intro:MM}Mirror Matter}

Mirror matter was originally discussed by Lee and Yang~\cite{Lee1956} in 1956 as an attempt to preserve parity as an unbroken symmetry of Nature after their discovery of parity violation in the weak interaction. They suggested that the transformation in the particle space corresponding to the space inversion \textbf{x}$\to -$\textbf{x} was not the usual transformation P but PR, where R corresponds to the transformation of a particle into a reflected state in the mirror particle space.

The idea was further developed by A.~Salam~\cite{Salam1957} and was clearly formulated in 1966 as a concept of the mirror universe by Kobzarev, Okun and Pomeranchuk~\cite{Kobzarev1966}. They proposed a model in which mirror and ordinary matter communicate predominantly through gravity. This concept evolved further into two versions, the symmetric (developed by Foot, Lew and Volkas~\cite{Foot1991}), and the asymmetric (proposed by Berezhiani and Mohapatra~\cite{Berezhiani1995}). For further historical details, see the review by Okun~\cite{Okun2006}.

The symmetric model provides a viable experimental signature through positronium (Ps). The main idea is that each ordinary particle (i.e.\ photon or electron) has a mirror particle with the same properties (e.g.\ mass and charge) but opposite chirality. These mirror particles would be singlets under the standard $\text{G}\equiv\text{SU}\left(\num{3}\right)_\text{C}\otimes\text{SU}\left(\num{2}\right)_\text{L}\otimes\text{U}\left(\num{1}\right)_\text{Y}$ gauge interactions~\cite{Foot1991}. Interactions within mirror particles are identical to their mirror partners: mirror electron and mirror photon will interact with each other in the same way ordinary electron and ordinary photon do. Having opposite chirality, parity conservation is restored at a global level.

Being massive and stable, mirror particles are a very good candidate for Dark Matter, because they interact with ordinary matter primarily through gravitation~\cite{Blinnikov1982, Blinnikov1983, Berezhiani2001, Berezhiani2005, Blinnikov2010}. However, the model allows other interactions, limited by charge conservation in each sector. Neutral particles and composites can in principle mix with their respective mirror partner, e.g.\ neutrinos~\cite{Foot1999}, photons~\cite{Foot2001}, the neutral Higgs boson~\cite{Barbieri2005,Ignatiev2000}, neutrons~\cite{Berezhiani2006,Ban2007,Serebrov2007,Pokotilovski2006} or muonium~\cite{Gninenko2013}. The photon - mirror photon ($\gamma$ - $\gamma\vtick$) mixing mechanism would then induce the Ps - Ps\vtick oscillation through the one-photon virtual annihilation channel of ortho-positronium.

Mirror matter can also provide a natural explanation for the similarity between Dark Matter and ordinary baryonic fractions, $\Omega_\text{DM}\simeq\SI{5}{\Omega_\text{b}}$. Although it is true that ordinary and mirror matter would have the same microphysics, that does not necessarily imply they should follow identical cosmological realizations. As pointed out by Berezhiani et al.~\cite{Berezhiani2001}, one can assume that the inflationary reheating temperature of the mirror sector $T\text{\vtick}$ was lower than the ordinary one $T$. With this premise, and since the two sectors can only interact very weakly, they would not reach thermal equilibrium with each other in early stages of the universe and hence would evolve independently during the Universe expansion. Moreover, since baryonic asymmetry (BA) depends on the departure from thermal equilibrium, it is possible that the BA is larger in the mirror sector than in the ordinary one. A temperature ratio $T\text{\vtick}/T<\num{0.2}$ could lead to mirror baryonic densities $\num{1}\leq\Omega_\text{DM}/\Omega_\text{b}\leq\num{5}$ compatible with the latest Planck Mission results~\cite{Planck2015}.

Finally, Mirror Matter is also an excellent candidate to explain the annual modulation reported by the DAMA Collaboration over \num{14} annual cycles with the former DAMA/NaI experiment and with the second generation DAMA/LIBRA phase1 (see~\cite{Bernabei2016} and references therein for latest reviews). The observed modulation has a period $T=\SI{0.998(2)}{years}$ and a phase $t_0=\SI{144(7)}{days}$, in good agreement with expectations for a Dark Matter annual modulation signal. These results give model-independent evidence for the presence of Dark Matter particles in the galactic halo, at a $\SI{9.3}{\sigma}$ C.~L.

Cerulli et al.~\cite{Cerulli2017} recently showed how the mirror sector can successfully describe such modulation, providing detailed characterizations of several chemical compositions of the mirror sector compatible with cosmological bounds. In particular, for a reference DM density of $\rho_\text{DM} = \SI{0.3}{\giga\electronvolt\per\cubic\centi\meter}$ and many different halo temperatures and compositions, they calculate coupling constants in the region $\varepsilon\sim\num{E-9}$, which has not been ruled out by any cosmological limits or direct experimental measurements. For comparison, the upper limit deduced  by the successful prediction of the primordial He$^{\num{4}}$ abundance by the SM is~\cite{Carlson1987}:
\begin{equation}
	\varepsilon\leq\num{3E-8}
	\label{eq:epsilonLimitBBN}
\end{equation}

\subsection{\label{subsec:Intro:PsPortal}Positronium as a Portal into the Mirror World}
Mirror and ordinary particles interact with each other predominantly by gravity. However, in \num{1986} Holdom~\cite{Holdom1986} pointed out that any new particle gauged by a new $\text{U}\left(\num{1}\right)$ would couple with a certain constant $\varepsilon$, thus effectively providing fractional charge to the new particles.

Glashow~\cite{Glashow1986} realized that this coupling would lead to a kinetic mixing of photons and mirror photons, described by the interaction Lagrangian density:
\begin{equation}
	L = \varepsilon F^{\mu\nu} F\vtick_{\mu\nu}
	\label{eq:LagrangianMirrorMatter}
\end{equation}
where $F^{\mu \nu}$ and $F\vtick_{\mu \nu}$ are the field strength tensors for electromagnetism and mirror electromagnetism respectively.

Due to its one-photon virtual annihilation channel, ortho-positronium and mirror ortho-positronium are connected and the degeneracy between the mass eigenstates is broken~\cite{Glashow1986}. The vacuum eigenstates
\begin{equation*}
	\frac{\ops + \opsM}{\sqrt{\num{2}}}
	\quad\quad
	\frac{\ops - \opsM}{\sqrt{\num{2}}}
	\label{eq:oPsEigenstates}
\end{equation*}
are therefore split in energy by $\Delta E = \num{2}h\varepsilon f$, where $f = \SI{8.7E4}{\mega\hertz}$ is the contribution to the ortho -- para splitting from the one-photon virtual annihilation diagram. This splitting leads to Rabi oscillations in which a state that is initially ordinary ortho-positronium will oscillate into its mirror partner with a probability
\begin{equation}
	\mathcal{P}\left(\ops\to\opsM\right) = \sin^{\num{2}}\Omega t
	\label{eq:RabiOscillation}
\end{equation}
where $\Omega = \num{2}\pi f\varepsilon$ is the oscillation frequency. Mirror matter having the same micro-physics as ordinary matter, \opsM\ will decay into mirror photons, which are very weakly coupled to ordinary matter and thus not detected. Such oscillations will therefore result in an apparent $\ops\to\text{invisible}$ process with a branching ratio
\begin{equation}
	\text{Br}\left(\ops\to\text{invisible}\right) = \frac{2\Omega^2}{\Gamma^2_\text{SM} + 4\Omega^2}
	\label{eq:BrRatioInVacuum}
\end{equation}
where $\Gamma_\text{SM}$ is the Standard Model decay rate of \ops~\cite{Adkins2000,Vallery2003}. Assuming $\varepsilon=\num{4E-9}$, the oscillation probability is
\begin{equation}
	\text{Br}\left(\ops\to\text{invisible}\right) = \num{2E-7}
	\label{eq:BrRatioIdeal}
\end{equation}
Note that, within the SM, photonless (and thus invisible) decays of both \ops\ and \pps\ into neutrinos are mediated by the weak interaction and are heavily suppressed with a branching ratio below \num{E-17} due to the small mass of the positronium atom~\cite{Pokraka2016,Czarnecki1999}.

The above calculations do not consider incoherent processes (e.g.~collisions with matter) and the effect of energy shifts induced by electromagnetic fields. Using the density matrix approach, the effect of electromagnetic fields on the branching ratio is shown to be negligible within the region of interest for this experimental search ($E\sim\SI{10}{\kilo\volt\per\cm}$ and $B\sim\SI{80}{\gauss}$)~\cite{Crivelli2010}. On the other hand, collisions of \ops\ with matter play a major role and can be source of large systematic effects and uncertainties~\cite{Demidov2011}.

In general, the number of collisions per lifetime will not be a well-defined value but rather a discrete distribution, i.e.~a certain fraction of the total \ops\ population $f_n$ will undergo $n$ collisions per lifetime with the corresponding branching ratio $\mathit{Br}_n$. One can thus calculate the total branching ratio as the weighted average
\begin{equation}
	\displaystyle
	\mathit{Br} = \sum\limits_{n=0}^\infty f_n\cdot\mathit{Br}_n
	\label{eq:WeightedAverageBrRatio}
\end{equation}

\subsection{\label{subsec:Intro:Limit}Current Experimental Limit on Br\texorpdfstring{$\left(\ops\to\text{invisible}\right)$}{(o-Ps --> invisible)}}

Ortho-positronium is a sensitive probe to test Mirror Matter models with two possible experimental signatures, namely the missing energy of the expected SM \SI{1.022}{\mega\electronvolt} decay or an apparent excess in the \ops\ decay rate compared to QED predictions~\cite{Gninenko1994}.

Previous measurements of missing energy in \ops\ decays were performed in the presence of matter~\cite{Atoyan1989,Mitsui1993,Badertscher2007}, where very high collision rates ($N\sim\num{E5}$) are expected, resulting in large uncertainties and strong suppression of the oscillation probability. Regarding discrepancies between QED predictions and experimental measurements of the decay rate, the most accurate measurements are still very far from QED precision~\cite{Adkins2000,Vallery2003}. Although these experiments are performed in a vacuum cavity with low collision rates, the lifetime calculation requires extrapolations to account precisely for the \emph{disappearance} of \ops\ into regions of lower gamma detection efficiency. It is therefore possible that any contribution $\Gamma_{\ops}^\text{inv.}$ could be artificially corrected by this extrapolation.

Both signatures have provided so far limits for $\varepsilon$ in the range \num{E-6} to \num{E-7} but suffer from large uncertainties. It is thus evident that an experiment with low collision rates (hence in vacuum) but without the need of any extrapolation (hence the missing energy technique) is necessary to confront Mirror Matter as a candidate to explain the DAMA/LIBRA anomaly.


\section{\label{sec:ExpSetup}Experimental Setup}

The working principle of our experiment is a vacuum cavity where ortho-positronium is confined, surrounded by a hermetic calorimeter to detect the photons expected for a SM decay. The resulting energy spectrum is centered at \SI{1.022}{\mega\electronvolt}, with a tail due to energy losses and inefficiencies extending down to low energies.

In case of $\ops\to\opsM$ oscillation, the experimental signature would be the absence of this energy deposition in the calorimeter. From simulation, it is possible to estimate the experimental background as the probability to misidentify an actual \ops\ decay (or in general a positron annihilation) as a zero-energy event. Therefore, if oscillations $\ops\to\opsM$ occur, an excess over the simulation prediction of such zero-energy events would be detected.

Note that \ops\ confined in a vacuum cavity will undergo collisions with the walls, whose rate can be modulated by tuning the \ops\ kinetic energy. A larger collision rate will suppress the $\ops\to\opsM$ oscillation probability while keeping the background constant. A possible signal observation can thus be cross-checked with controlled collision rate modulation~\cite{Crivelli2010}.

\subsection{\label{subsec:ExpSetup:Beam}The Slow Positron Beam}

The slow positron beam at ETHZ is based on a \SI{120}{\mega\becquerel} $^{22}$Na radioactive source coupled to a tungsten mesh acting as a moderator providing a flux of $\Phi_{\si{\positron}}^\text{slow}\sim\SI{15000}{\positron\per\second}$. Slow ($<\SI{3}{\electronvolt}$) positrons are electrostatically accelerated to \SI{200}{\electronvolt} and magnetically guided with a set of Helmholtz coils which creates a radially confining field. A high-efficiency tagging system is used to detect the arrival of a positron to the cavity where the positronium converter is placed. The beam is equipped with a velocity selector and a bunching system based on drift tubes and grids where time-dependant potentials are applied~\cite{Alberola2006}.

\subsubsection{\label{subsubsec:ExpSetup:Beam:Tagging}The Positron Tagging System}

\begin{figure}
	\centering
	\includegraphics[width=\linewidth, page=2]{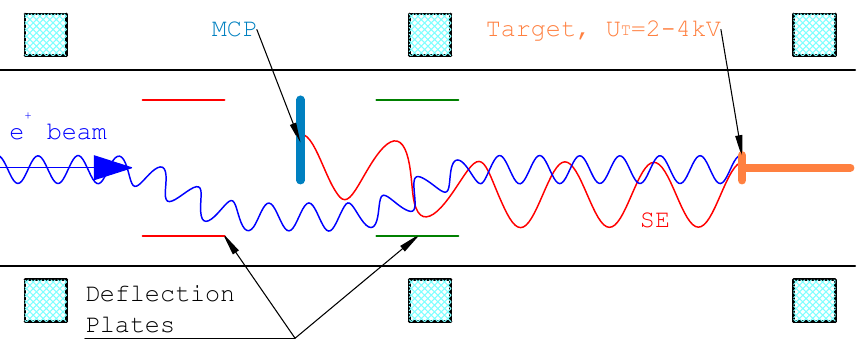}
	\caption{\label{fig:ExpSetup:SketchTagging}Positron tagging scheme with a Micro-Channel Plate. The positron beam (blue helix, coming from the left) is deflected off axis by the deflection plates (red and green) to bypass the micro-channel plate (MCP). Secondary Electrons (SE, red helix) are released when the positron impinges the target and guided back and detected by the MCP. Note that the positron and electron trajectories are only sketches, the actual deflection is perpendicular to the drawing plane.}
\end{figure}

The tagging of positrons is a crucial feature of the experimental setup. It is used to define the arrival of a positron to the target and therefore serves as the \textit{START} signal for the DAQ. The tagging system is based on a Micro-Channel Plate (MCP) which detects Secondary Electrons (SE) released by a positron hitting the target (Fig.~\ref{fig:ExpSetup:SketchTagging}).

The positron beam, transversally confined by the magnetic field, is slightly deflected off axis by deflection plates (red), which effectively work as an $E\times B$ filter. Positrons can thus bypass the MCP and are deflected back on axis by an opposite pair of deflection plates (green) and continue their way downstream to the target. They are then accelerated by the target potential $U_\text{T}$, where they are implanted and release SE. These electrons are accelerated backwards by the same $U_\text{T}$ and guided by the same magnetic field. However, when reaching the deflection plates, the electrons are deflected towards the MCP, where they are detected.

This tagging systems presents two important limitations, namely dark counts and detection of other charged particles. Even though the MCP was specifically selected for its low dark counts rate (Hamamatsu F4655-12), it is still at the level of \SI{1}{\hertz}. These accidentals are uniformly distributed in time and uncorrelated with the arrival of a positron into the target, and are therefore a source of background. Regarding the detection of other charged particles, it was found that some positrons may annihilate close to or even against the MCP due to transportation inefficiencies. These positrons, or the SE following them, may be detected by the MCP, leading to a trigger accidental correlated with the positron flux, but not with the presence of a positron inside the calorimeter.

\subsubsection{\label{subsubsec:ExpSetup:Beam:Chopper}The Chopper}
The chopper is a grid placed in front of the tungsten moderator and set above the moderator potential $U_\text{M} = \SI{200}{\volt}$ to constantly block the low energy positrons. Driven by a global clock, the chopper is pulsed down below $U_\text{M}$, thus letting through the positrons during a time window $t_\text{W} = \SI{300}{\nano\second}$, at a frequency $f = \SI{333}{\kilo\hertz}$ ($T = \SI{3}{\micro\second}$). The chopping system suppresses positron pile-up, i.e.~the presence of more than one positron in the cavity within the same event, which introduces a signal inefficiency. The total signal efficiency can be measured using a trigger uncorrelated with the positron beam and was found to be $\eta_\text{S} = \SI{92.1}{\percent}$.

\subsubsection{\label{subsubsec:ExpSetup:Beam:Buncher}The Buncher}

\begin{figure}
	\centering
	\includegraphics[width=\linewidth, page=1]{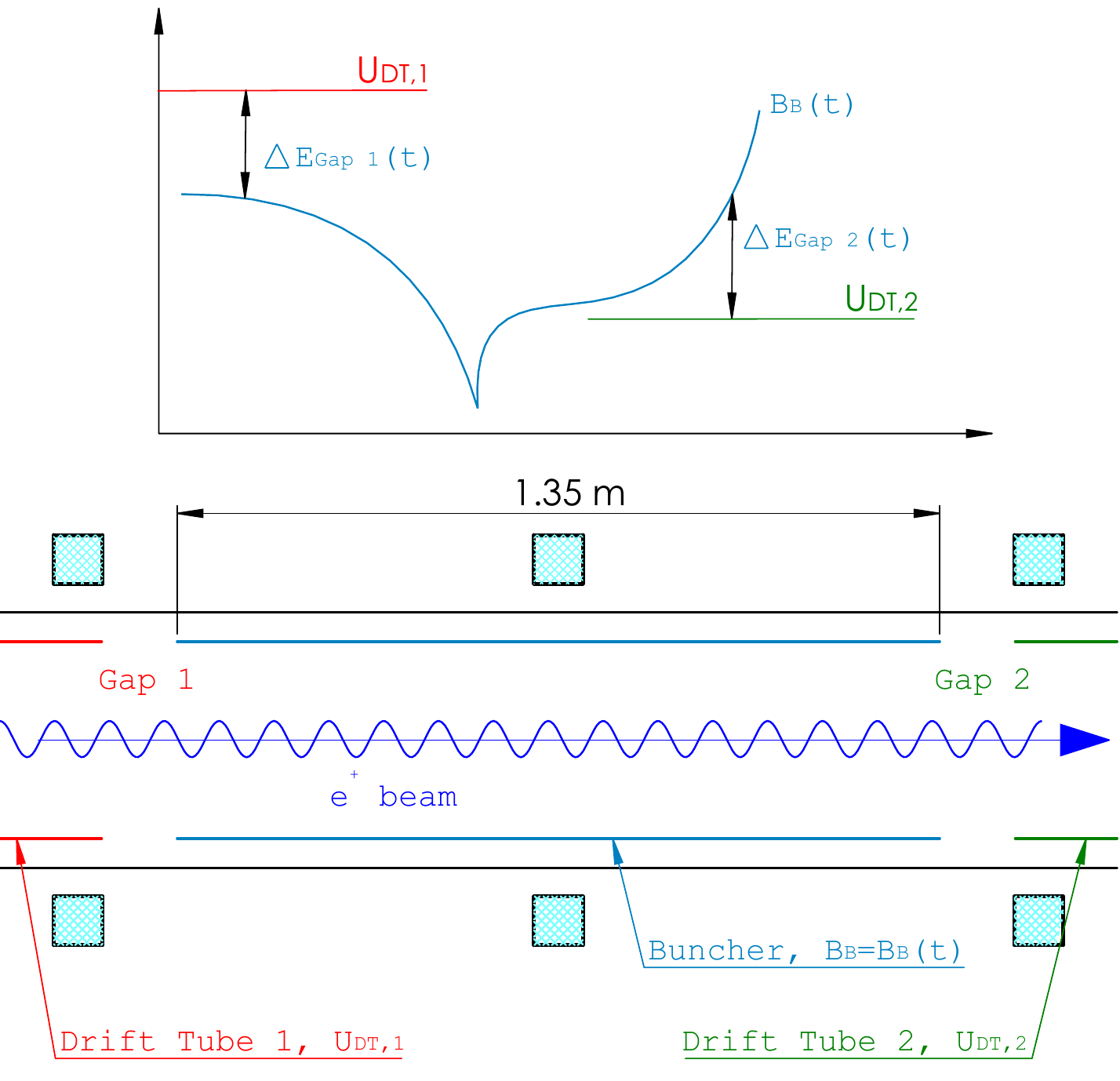}
	\caption{\label{fig:ExpSetup:SketchBunching}Positron bunching scheme.}
\end{figure}

The \SI{300}{\nano\second} positron pulse is compressed into few \si{\nano\second} by means of a double-gap buncher~\cite{Alberola2006} sketched in Fig.~\ref{fig:ExpSetup:SketchBunching} to increase the signal-to-noise ratio and reduce triggering-related background, e.g.~MCP noise. A positron arriving at Gap~1 is accelerated by the time-dependent potential difference so that late positrons will acquire a larger velocity. If the potential in the buncher is set properly, a linear velocity distribution can be achieved, producing an ideal compression into a shorter positron bunch. The same process is repeated at Gap~2, the goal being to compress the bunch as much as possible when it reaches the target. 

\begin{figure}
	\centering
	\includegraphics[width=\linewidth]{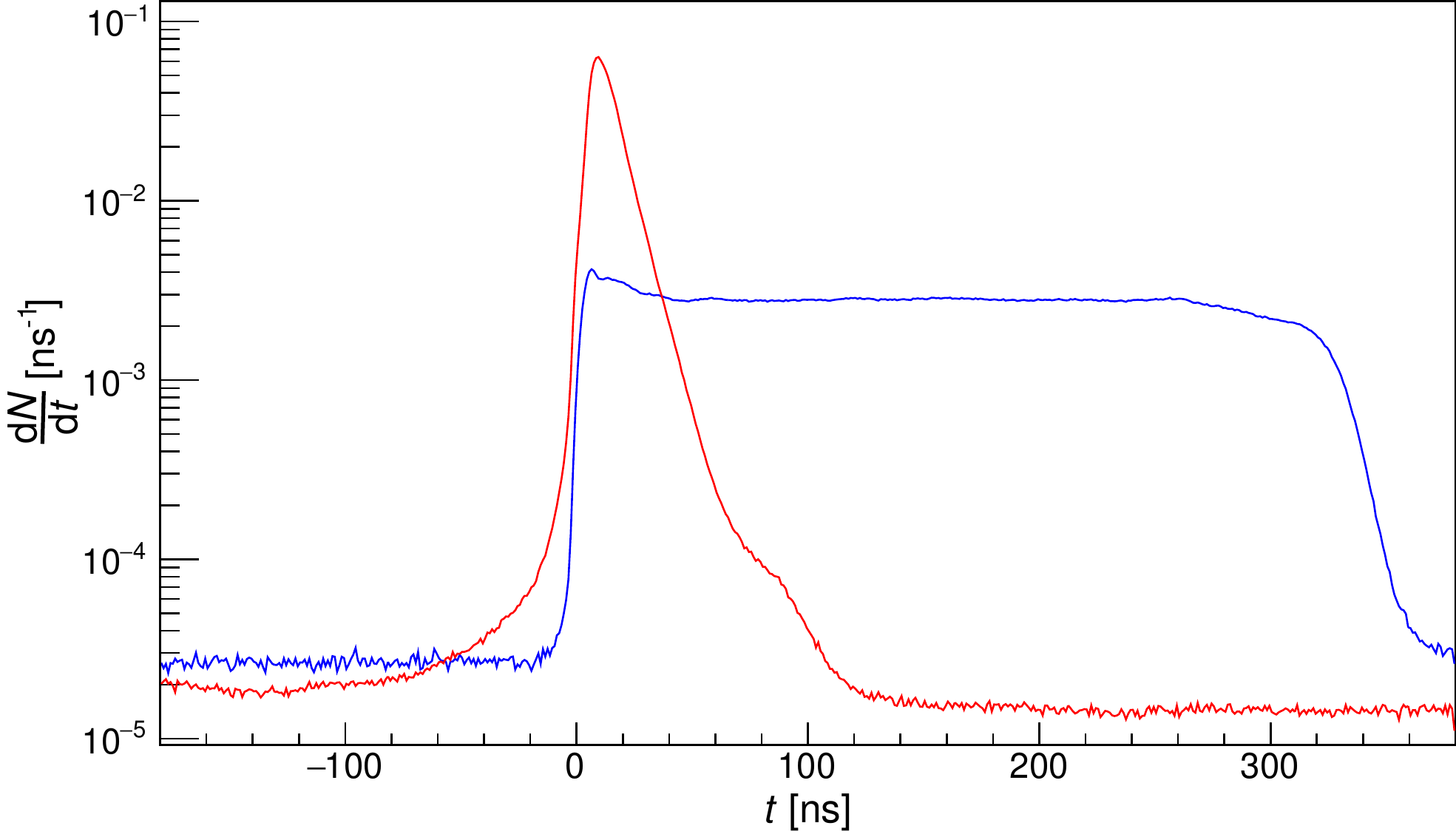}
	\caption{\label{fig:ExpSetup:Bunching}Time distribution of unbunched (blue) and bunched (red) positrons on the target.}
\end{figure}

The initial \SI{300}{\nano\second} bunches are compressed to \SI{14}{\nano\second} FWHM pulses (Fig.~\ref{fig:ExpSetup:Bunching}). The energy range of the positrons is given by
\begin{equation}
	U_\text{M}-U_\text{B} < E_{\si{\positron}} < U_\text{M}+U_\text{B}
	\label{eq:ExpSetup:PosEnergyDist}
\end{equation}
where $U_\text{B} = \SI{60}{\volt}$ and $U_\text{M} = \SI{200}{\volt}$ are the buncher amplitude and the moderator potential.

\subsubsection{\label{subsubsec:ExpSetup:Beam:ReImplElectrode}The Re-implantation Electrode}

When positrons reach the positronium cavity, they are accelerated to few \si{\kilo\electronvolt} and impinge on the target. They quickly slow down and diffuse in the bulk, where they will generally pick-up an electron. However, a positron can also reach the surface again before losing all its kinetic energy and escape into vacuum, i.e.~it can be backscattered~\cite{Crivelli2010}. The energy range of these positrons is
\begin{equation}
	E_{\si{\positron}}^\text{back.}\in\left[0, U_\text{T}+U_\text{M}+U_\text{B}\right]
	\label{eq:ExpSetup:BackPosEnergyDist}
\end{equation}
where $U_\text{T}$ is the target potential and $U_\text{M}+U_\text{B}$ is the maximum initial energy from Eq.~\eqref{eq:ExpSetup:PosEnergyDist}. Backscattered positrons with enough energy to escape the cavity ($E_{\si{\positron}}^\text{back.} > U_\text{T}$) are expected at a level \num{E-4} (see Table~\ref{tab:ExpSetup:Background:BackPos}), becoming a source of background.

A re-implantation electrode is placed at the end of the beam line, before the cavity, to ensure no backscattered positron escapes the cavity. The electrode is set to a low potential, below the minimum positron energy $U_\text{M}-U_\text{B}$. After the compressed bunch of positrons has passed through, the electrode potential is raised above the maximum positron energy, $U_\text{M}+U_\text{B}$, blocking all backscattered positrons and reducing the background below \num{E-6} (see Table~\ref{tab:ExpSetup:Background:BackPos}).

\subsection{\label{subsec:ExpSetup:Cavity}The Ortho-Positronium Cavity}
\subsubsection{\label{subsubsec:ExpSetup:Cavity:Production}Ortho-Positronium Production in Vacuum}
Positronium production in vacuum is achieved with a porous film where a positron impinges and captures an electron from the bulk to form \ops, which diffuses through the porous structure back into vacuum. Different samples that can be used to produce positronium were studied and characterized~\cite{Liszkay2008,Liszkay2009,Cassidy2010,Crivelli2010_2}, the most promising positronium converters for this experiment being thin silica films prepared on a rigid substrate with a non-ionic surfactant, which is later removed by heating at \SI{450}{\celsius} to obtain the porous structure.

The samples were prepared on a \SI{110}{\micro\meter} thick, \SI{15}{\milli\meter} diameter borosilicate disc with a \SI{10}{\nano\meter} layer of gold deposited in the opposite face of the disc to improve conductivity. This thin substrate reduces photon energy losses which could lead to background. The main features of these films (see Table~\ref{tab:ExpSetup:OPsProperties}) are a high and constant yield of \ops\ in vacuum ($y_{\ops}\sim\SI{30}{\percent}$) for the implantation energy interval \SIrange{2}{4}{\kilo\electronvolt} and an \ops\ re-emission energy dependent on positron implantation energy~\cite{Crivelli2010_2}.

\begin{table}
	\centering
	\caption{Ortho-positronium yield $y_{\ops}$ and mean kinetic energy $E_{\ops}$ from Ref.~\cite{Crivelli2010_2} and estimated average number of collisions per lifetime $N_\text{coll}$ from simulation, for different positron implantation energies $E_{\si{\positron}}$. The branching ratio $\text{Br}\left(\ops\to\opsM\right)$ is calculated according to Eq.~\eqref{eq:WeightedAverageBrRatio}.}
	\label{tab:ExpSetup:OPsProperties}
	\begin{ruledtabular}
	\begin{tabular}{c c c c c}
		$E_{\si{\positron}}$ $\left[\si{\kilo\electronvolt}\right]$
			& $y_{\ops}$
			& $E_{\ops}$ $\left[\si{\milli\electronvolt}\right]$
			& $N_\text{coll}$
			& $\text{Br}\left(\ops\to\opsM\right)$
		\\
		\hline
		2 & \SI{30}{\percent} & 440	& 3.37 & \num{1.1E-7} \\
		3 & \SI{30}{\percent} & 220	& 2.42 & \num{9.8E-8} \\
		4 & \SI{29}{\percent} & 130	& 1.87 & \num{8.3E-8} \\
	\end{tabular}
	\end{ruledtabular}
\end{table}

The positronium cavity consists of a \SI{0.7}{\milli\meter} thick, \SI{17}{\milli\meter} diameter aluminum pipe where the positronium converter is attached and set to a potential $U_\text{T}$ to accelerate incoming positrons. A thin aluminum wire (core diameter of \SI{200}{\micro\meter}) was coiled around the cavity to create a homogeneous magnetic field to guide positrons into the cavity and extract secondary electrons.

\subsubsection{\label{subsubsec:ExpSetup:Cavity:Modulation}Signal Modulation}

\begin{figure}
	\centering
	\includegraphics[width=\linewidth, page=1]{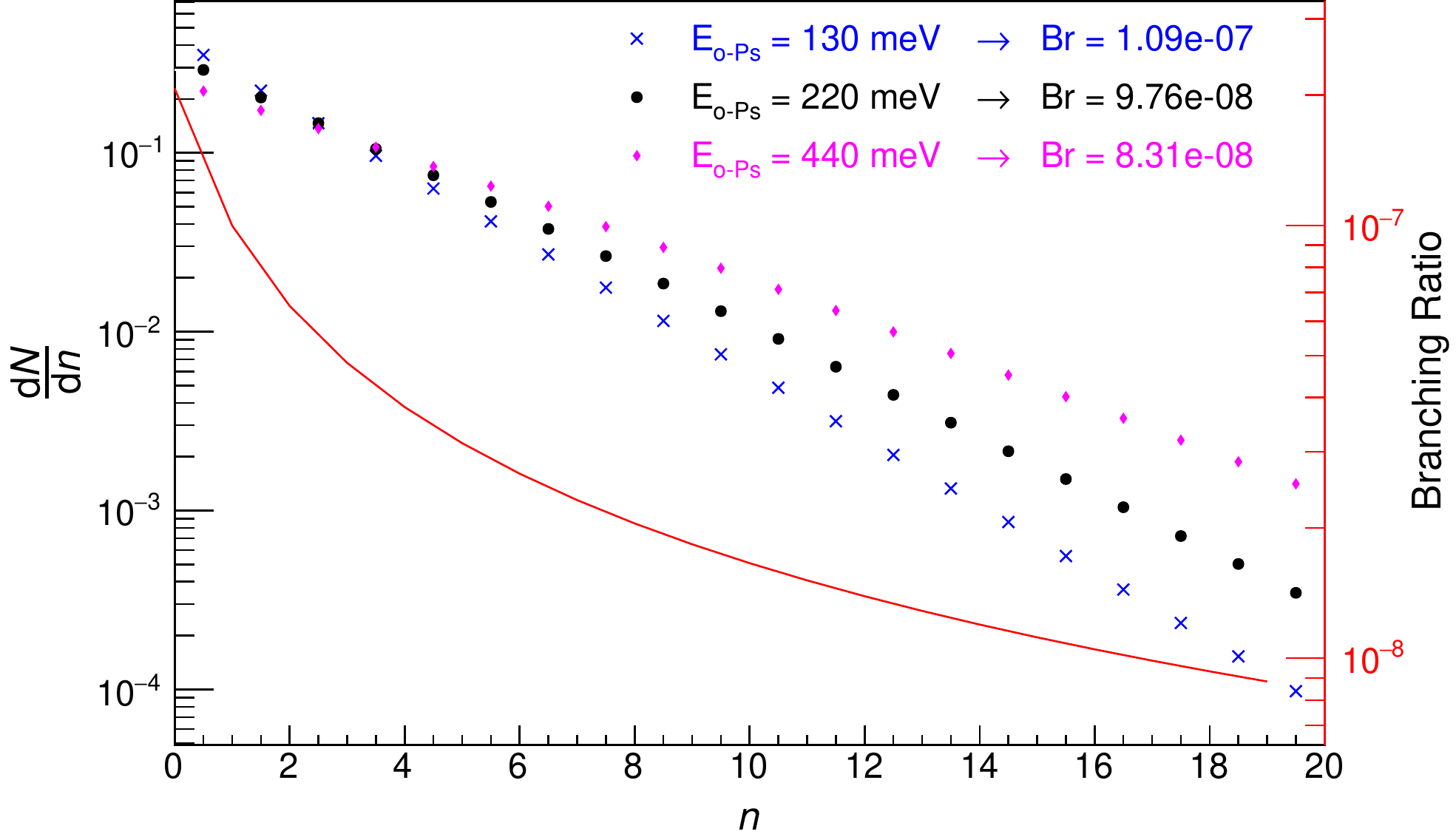}
	\caption{\label{fig:ExpSetup:oPsCollisions}Distribution of events with total number of \ops\ collisions per lifetime $n$ for different \ops\ emission energies, from Geant4 simulation. Red solid line shows the branching ratio for the process $\ops\to\opsM$ for a coupling constant $\varepsilon=\num{4E-9}$, electric field $E = \SI{10}{\kilo\volt\per\cm}$ and magnetic field $B = \SI{70}{\gauss}$. Dotted points are frequencies of events with $n$ collisions from simulations with different \ops\ emission energies. The averaged branching ratios are calculated according to Eq.~\eqref{eq:WeightedAverageBrRatio} for each monoenergetic case.}
\end{figure}

In case of signal observation, a key feature of the experiment is the possibility to check that the origin of the signal is due to new physics rather than an underestimation of the background. Due to the signal suppression induced by collisions of \ops\ with matter, one can modulate the signal by tuning the velocity of \ops\ and thus the rate of collisions~\cite{Crivelli2010}. Figure~\ref{fig:ExpSetup:oPsCollisions} shows the distribution of events with different number of total collisions per lifetime at different positronium kinetic energies, corresponding to the implantation energies \SIlist{4;3;2}{\kilo\electronvolt} from Table~\ref{tab:ExpSetup:OPsProperties}. As expected, more energetic positronium (from a shallower implanted positron) shows a larger frequency of many-collisions events. As a guide for the eye, the branching ratio for the different collision rates is also plotted, assuming $\varepsilon=\num{4E-9}$. Note that events with many collisions with the cavity have a much lower probability to oscillate into \opsM.

For each \ops\ kinetic energy, one can then calculate the average branching ratio with Eq.~\eqref{eq:WeightedAverageBrRatio}. The difference between such branching ratios for two given \ops\ energies provides the modulation of the oscillation signal.

\subsection{\label{subsec:ExpSetup:ECAL}The Calorimeter}

\begin{figure*}
	\centering
	\def\svgwidth{\linewidth}
	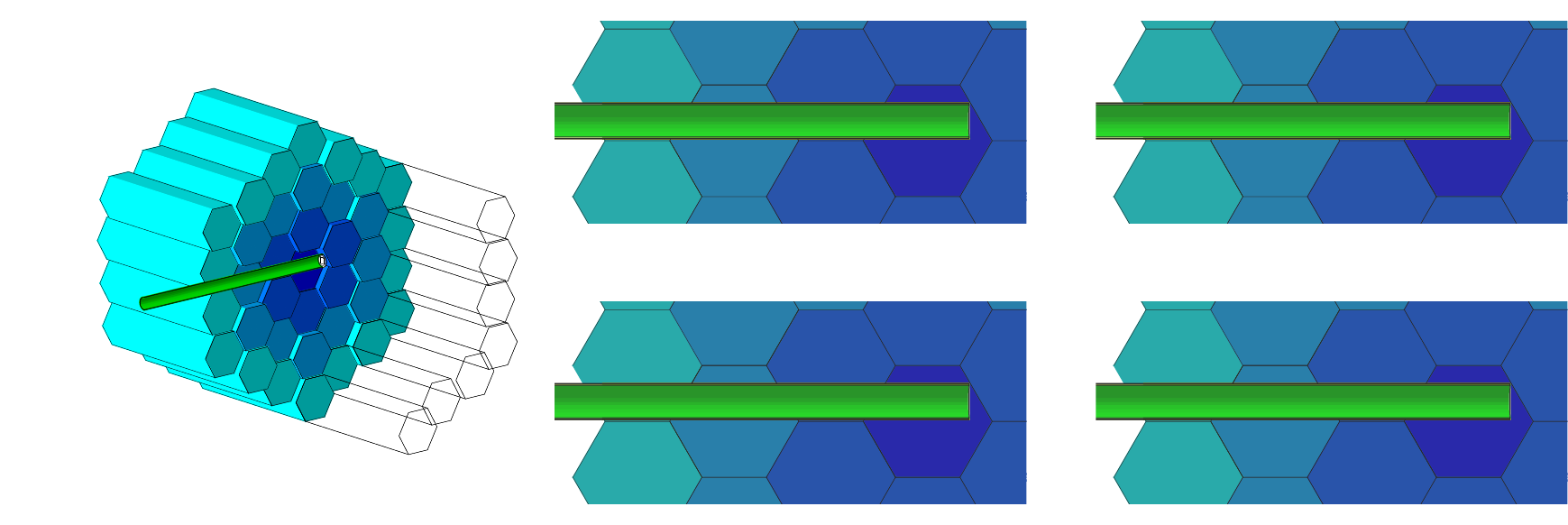
	\caption{\label{fig:ExpSetup:ECAL}\emph{Left}: sectional view of the calorimeter with the vacuum cavity and the positronium converter. Some scintillators from the cut side are shown as wireframes for reference.\\
	\emph{Right}: sketches of possible events\\
	\ \ \ $1)$ prompt annihilation and \pps\ decay into two back-to-back \SI{511}{\kilo\electronvolt} photons.\\
	$2)$ \ops\ decay into three photons.\\
	$3)$ $\ops\to\opsM$ oscillation, \opsM\ then decays into three \gammaM\ which are not detected in the ECAL.\\
	$4)$ background event where a photon is absorbed in the cavity and two photons escape the ECAL.}
\end{figure*}

The experimental signature of $\ops\to\opsM$ is the absence of energy deposition in a hermetic calorimeter (ECAL) surrounding the \ops\ cavity. The calorimeter consists of \num{92} BGO (Bi$_{12}$GeO$_{20}$) scintillators placed in a two-halves honeycomb structure with an aperture to accommodate the positronium cavity (Fig.~\ref{fig:ExpSetup:ECAL}). Each individual detector is a scintillating crystal (a \SI{200}{\milli\meter} long and \SI{55}{\milli\meter} wide hexagonal prism) and a thin wrapping around the crystal to increase light collection, improve energy resolution and reduce cross-talk between neighbouring detectors. Each detector is coupled to a Photo-Multiplier Tube (PMT) which collects the scintillating light, producing an amplified electrical signal proportional to the original photon energy.

The detector is mounted inside a light-tight PVC black box to avoid the PMTs detecting natural light as a signal. Since the gains of both BGO scintillators and PMTs are very sensitive to temperature (effective temperature coefficient estimated to be \SI{-1.38}{\percent\per\degreeCelsius}~\cite{Wang2014}), the PVC box is equipped with two copper plates coupled to a temperature controlled water circuit, resulting in the long-term stability necessary for the data taking. To improve heat extraction from the cavity solenoid, pre-cooled pressurized air is fed into the ECAL, greatly increasing convection efficiency.

Each PMT signal is read out individually via a set of CAEN V792 QDC modules, which integrate the current over a time $t_\text{G} = \SI{3}{\micro\second}$ after the $t_\text{START}$ signal from the MCP. The probability of \ops\ decaying after $t_\text{G}$ is 
\begin{equation}
	\begin{aligned}
	S	&= \int\limits_{t_\text{G}}^\infty \frac{1}{\tau_{\ops}}\exp\left(-\frac{t}{\tau_{\ops}}\right)\dd{t} = \exp\left(-\frac{t_\text{G}}{\tau_{\ops}}\right) \\
		&= \num{6.7E-10}
	\end{aligned}
	\label{eq:DecayProbability}
\end{equation}
well below the expected sensitivity.

The $\ops\to\opsM$ signal is defined as the absence of energy deposition in any crystal. Due to finite energy resolution and contribution of electronics and PMT noise, one must set for each BGO $i$ a certain threshold $E_{\text{T},i}$ below which the energy deposition is considered to be zero. These thresholds were individually picked to minimize both signal inefficiency (i.e.~the probability to misidentifiy a zero-energy event as a SM decay) and energy losses (energy depositions in a single BGO below $E_{\text{T},i}$ will not be accounted for, leading to possible background).

The individual energy depositions are thus summed up to obtain the total energy $E_\text{ECAL}$
\begin{equation}
	E_\text{ECAL} = \sum\limits_{i}^{92} \begin{cases}
		0	& \mbox{if } E_i < E_{\text{T},i} \\ 
		E_i	& \mbox{if } E_i \geq E_{\text{T},i}
	\end{cases}
	\label{eq:SignalDefinition}
\end{equation}
Therefore the signal (zero-energy compatible events) is defined as those events with $E_\text{ECAL} = \num{0}$.

\subsection{\label{subsec:ExpSetup:Background}Background Sources}
The four following sources of background have been identified:

\subsubsection{\label{subsubsec:ExpSetup:Background:Accidentals}Trigger Accidentals}
Trigger accidentals happen when an MCP signal is detected without the presence of a positron in the cavity, and they represent the largest background contribution. Three different types were found:
\begin{itemize}
	\item MCP dark counts ($<\SI{1}{\hertz}$), which are uniformly distributed in time and unrelated to neither the positron beam nor the the implantation energy.
	\item Positron-related triggers. Due to beam transportation inefficiencies, some positrons may annihilate close to the MCP or even against it. The corresponding secondary electrons or even the annihilation photons can be detected by the MCP, leading to a time- and beam-dependent background.
	\item Electrons released from the positronium converter due to the strong electric field, which are transported upstream as if they were secondary electrons. This contribution is uniformly distributed in time but depends on the implantation energy.
\end{itemize}
Trigger accidentals are the dominant background ($>\num{E-4}$), but its rate can be experimentally determined, as will be shown later.

\subsubsection{\label{subsubsec:ExpSetup:Background:ECAL}Calorimeter}
The calorimeter was designed to ensure high hermeticity and minimize photon energy losses, verified by a detailed Geant4~\cite{Geant4} simulation of the complete setup. Energy depositions in dead material, i.e.~anything besides the scintillators, increase the probability to misidentify a positron annihilation as a zero-energy compatible event, especially when detector efficiency and energy resolution are taken into account. The simulation considers the contribution from all structural elements of the cavity (e.g.~the pipe, the solenoid or the \ops\ converter) as well as the scintillator wrappings. The kinetic energy of \ops\ is another key parameter: faster \ops\ is more likely to decay further upstream, where hermeticity decreases (see Fig.~\ref{fig:ExpSetup:ECAL}). With a kinetic energy $E_{\ops} = \SI{440}{\milli\electronvolt}$, corresponding to the shallowest implantation energy $E_{\si{\positron}} = \SI{2}{\kilo\electronvolt}$, the total background due to energy losses and hermeticity is at a level of \num{E-7}, below the experimental sensitivity.

\subsubsection{\label{subsubsec:ExpSetup:Background:BackPos}Backscattered Positrons}

As explained in Section~\ref{subsubsec:ExpSetup:Beam:ReImplElectrode}, positron backscattering is a very well known process which may lead to a tagged positron escaping the confinement cavity. A Geant4 simulation was used to obtain the positron backscattering fraction, as well as its energy and angular distribution, based on Ref.~\cite{Huang2015}. The trajectory of backscattered positron inside the vacuum pipe was then simulated with the beam optics package SIMION~\cite{SIMION}, reproducing the electric and magnetic fields in the vacuum cavity. Table~\ref{tab:ExpSetup:Background:BackPos} shows the simulated backscattering and escape probabilities at different implantation energies, which were at a level of \num{E-4} in very good agreement with measurements performed without the re-implantation electrode. This background can be suppressed below the experimental sensitivity with the use of the abovementioned re-implantation electrode.

\begin{table}
	\sisetup{separate-uncertainty=false}
	\centering
	\caption{\label{tab:ExpSetup:Background:BackPos}Simulated positron backscattering fraction and escape probabilities, with and without the re-implantation electrode, at different positron implantation energies $E_{\si{\positron}}$.}
	\begin{ruledtabular}
	\begin{tabular}{c c c c}
		$E_{\si{\positron}}$					& Backscattered			& \multicolumn{2}{c}{Background}		\\
		$\left[\si{\kilo\electronvolt}\right]$	& Fraction [\si{\percent}]	& Without electrode		& With electrode	\\[1pt]
		\hline
		2	& \num{5.861(7)}	& \num{1.79(4)E-4} & $<\num{4.5E-6}$ \\
		3	& \num{6.882(8)}	& \num{1.28(4)E-4} & $<\num{3.7E-6}$ \\
		4	& \num{7.484(9)}	& \num{1.02(3)E-4} & $<\num{4.1E-6}$ \\
	\end{tabular}
	\end{ruledtabular}
	\sisetup{separate-uncertainty=true}
\end{table}

\subsubsection{\label{subsubsec:ExpSetup:Background:FastOPs}Fast Backscattered \ops}

Ortho-positronium can be emitted from the converter with large kinetic energy (peaking around \SI{15}{\electronvolt}) due to backscattered positrons which capture an electron before exiting the surface~\cite{Crivelli2010}. Very energetic \ops\ is more likely to escape the high-efficiency detection volume. This possibility has been studied using a Geant4 simulation and similar assumptions from Ref.~\cite{Crivelli2010}: for \ops\ with kinetic energy below its binding energy (\SI{6.8}{\electronvolt}), the annihilation probability via pick-off when it collides with the pipe is very conservatively assumed to be zero, and \SI{100}{\percent} otherwise above the dissociation threshold~\cite{Gidley1995}.

Table~\ref{tab:ExpSetup:Background:FastOPs} shows the escape probability $\xi$ for some \ops\ kinetic energy. As expected, larger kinetic energies of \ops\ lead to a higher escape probability. However, note that above $E_{\si{\positron}} = \SI{6.8}{\electronvolt}$ the pick-off probability function changes from \num{0} to \SI{100}{\percent} and therefore the escaping probability is suppressed.

\begin{table}
	\sisetup{separate-uncertainty=false}
	\centering
	\caption{\label{tab:ExpSetup:Background:FastOPs}Simulated escape probabilities of fast backscattered \ops.}
	\begin{tabular*}{\columnwidth}{S[table-format=3.0(0)]@{\extracolsep{\fill}} S[table-format=1.3(1)e+1]}
		\toprule   \vspace{-6pt}\\
		{\ops\ energy $\left[\si{\electronvolt}\right]$} & {Escape Probability $\xi$} \\
		\hline
		3	& 3.392(4)E-4	\\
		6	& 1.238(3)E-3	\\
		7	& 5.4(1)E-5		\\
		20	& 1.17(2)E-4	\\
		100	& 2.17(2)E-4	\\
		\botrule
	\end{tabular*}
	\sisetup{separate-uncertainty=true}
\end{table}

This background estimation has to be integrated over the whole backscattered \ops\ energy spectrum, which can be assumed to be a Landau distribution peaked at \SI{15}{\electronvolt}, the maximum of the \si{\electron} capture cross section~\cite{Crivelli2010}. A rough estimation of the total escape probability $\xi_{\ops}$ gives $\xi_{\ops}<\num{E-4}$. The resulting background is then calculated as
\begin{equation}
	B_\ops = f_{\text{back. }\ops}\cdot\xi_{\ops}
	\label{eq:ExpSetup:Background:FastOPs}
\end{equation}
where $f_{\text{back. }\ops}$ is the fraction of incident positrons leading to fast backscattered \ops. This fraction decreases at larger positron implantation energies, and can be estimated to be $f_{\text{back. }\ops}<\SI{1}{\percent}$ already at $E_{\si{\positron}} = \SI{2}{\kilo\electronvolt}$~\cite{Gidley1995}, resulting in a background below \num{E-6}.


\section{\label{sec:Results}Results}
\begin{figure}
	\centering
	\subfloat[\label{fig:Results:DataSample:On}Target at \SI{1750}{\volt}.]{\includegraphics[width=\linewidth, page=1]{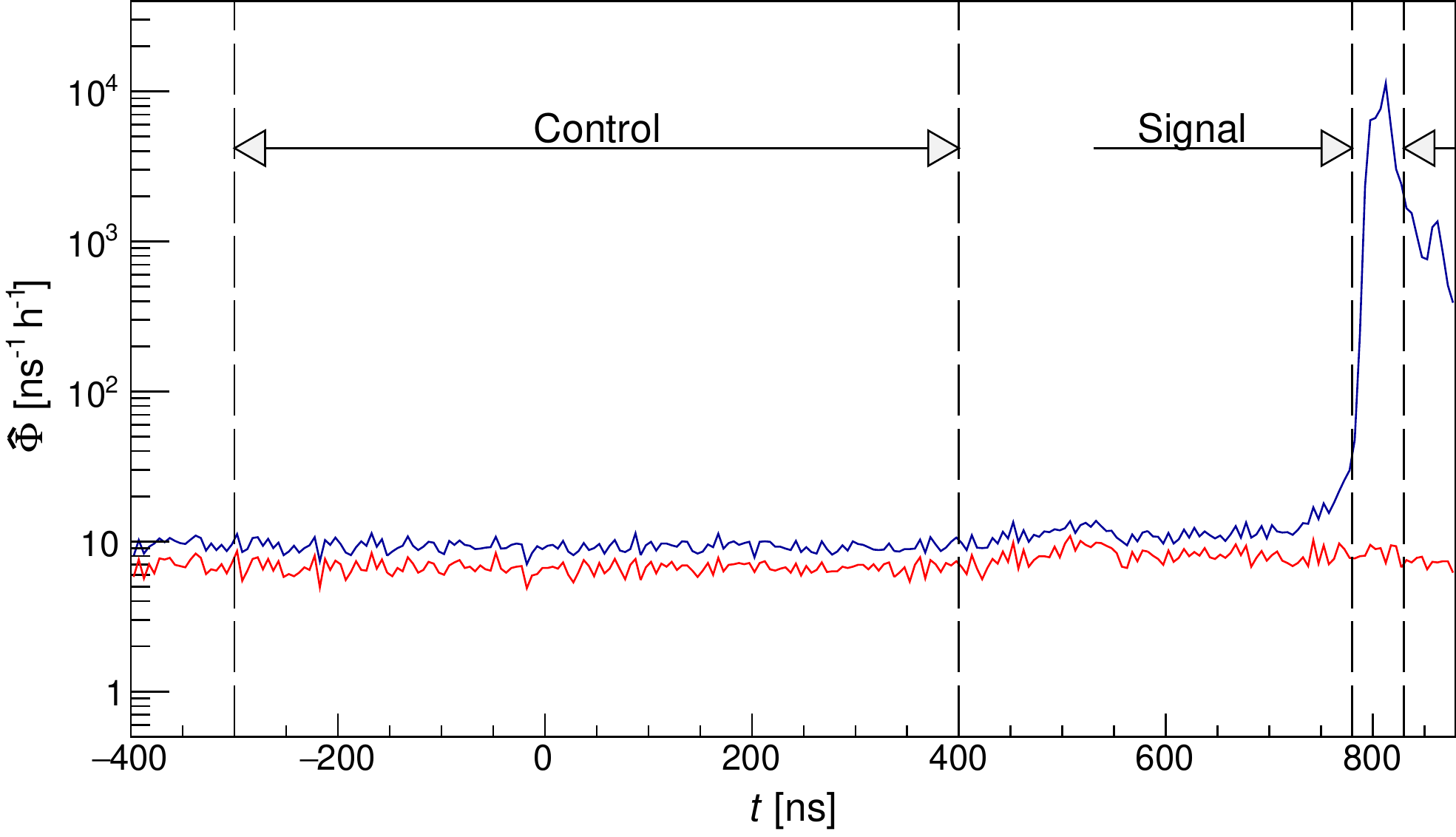}}

	\subfloat[\label{fig:Results:DataSample:Off}Target grounded.]{\includegraphics[width=\linewidth, page=2]{Results.pdf}}
	\caption{\label{fig:Results:DataSample}Time distribution of all (blue) and zero-energy (red) events is shown when the target is set at $U_\text{T} = \SI{1750}{\volt}$ (top) and when it is grounded (bottom).}
\end{figure}

Data were collected for positron implantation energies \SIrange{2}{4}{\kilo\electronvolt} for several days (see Table~\ref{tab:Results:DataSets}). Each data set consists of a collection of events, for which the event time $t$ (time difference between positron tagging and the chopper pulse measured with a CAEN V1290N Time-to-Digital Converter with $\SI{250}{\pico\second}$ resolution) and the energy depositions $E_i$ in every scintillator are recorded. The normalized event rate is defined as:
\begin{equation}
	\hat{\Phi} = \frac{1}{t_\text{acq}}\dv{N}{t}
	\label{eq:Results:NormalizedRateDefintion}
\end{equation}
where $t_\text{acq}$ is the acquisition time from Table~\ref{tab:Results:DataSets}. Figure~\ref{fig:Results:DataSample} shows the normalized event rate of all events (in blue) when the target is set to \SI{1750}{\volt} (top) and when it is grounded (bottom). The red line corresponds to the subset of zero-energy compatible events, i.e.~events with no energy deposition in any scintillator, $E_\text{ECAL} = 0$.

\begin{table}
	\centering
	\caption{\label{tab:Results:DataSets}Chronological relation of data sets, with positron implantation energy $E_{\si{\positron}}$ and acquisition times $t_\text{acq}$ (beam on) and $t^\text{b}_\text{acq}$ (beam off).}
	\begin{ruledtabular}
	\begin{tabular}{c c c c}
		Run ID
			& $E_{\si{\positron}}$ $\left[\si{\kilo\electronvolt}\right]$
			& $t_\text{acq}$ $\left[\si{\hour}\right]$
			& $t^\text{b}_\text{acq}$ $\left[\si{\hour}\right]$
		\\
		\hline
		1	& 2.0	& 13.6	& 7.9 \\
		2	& 3.0	& 16.7	& 11.8 \\
		3	& 4.0	& 18.2	& 18.7 \\
		4	& 3.5	& 10.9	& 26.9 \\
		5	& 3.0	& 41.9	& 5.4
	\end{tabular}
	\end{ruledtabular}
\end{table}

The grounded target configuration provides an excellent background estimation, because secondary electrons released by the incoming positrons are not accelerated and therefore do not reach the MCP. The small fraction of events with $E_\text{ECAL} > 0$ (difference between blue and red lines) is due to inefficiencies in the calorimeter, e.g.~cosmic rays and electronics noise, which were measured to be $\sim\SI{10}{\percent}$. When a negative potential is applied to the target, positrons are tagged and a clear excess of total events is observed around $t=\SI{815}{\nano\second}$. The signal is thus an excess of zero-energy compatible events within the same region.

\subsection{\label{subsec:Results:BackgroundEstimation}Background Estimation}

The background estimation provides the number of zero-energy compatible events $N_\text{B}$ to be expected in the signal region for each measurement. The background rate can be calculated as the combination of three normalized rates obtained in the control and signal regions defined in Fig.~\ref{fig:Results:DataSample}. One can then define, according to Eq.~\eqref{eq:Results:NormalizedRateDefintion}, the following mean normalized rates:
\begin{itemize}
	\item Target ON, control region: $\hat{\Phi}_{\text{B}, 1}$
	\item Target OFF, control region: $\hat{\Phi}_{\text{B}, 2}$
	\item Target OFF, signal region: $\hat{\Phi}_{\text{B}, 3}$
\end{itemize}
$\hat{\Phi}_{\text{B}, 1}$ accounts for all background contributions that are uniformly distributed in time, e.g.~MCP dark counts and electrons released due to the applied target potential $U_\text{T}$. When the target is off, the contribution from positrons annihilating at the MCP or its vicinity and being detected is not modified, which can be therefore estimated by the difference $\hat{\Phi}_{\text{B}, 3} - \hat{\Phi}_{\text{B}, 2}$. This value was found to depend on the beam configuration used to guide the positrons to the target, which has to be adjusted for each target potential. It was thus necessary to take a background measurement for each implantation energy. The expected background rate at the signal region with target ON is thus:
\begin{equation}
	\hat{\Phi}_\text{B} = \hat{\Phi}_{\text{B}, 1} + \hat{\Phi}_{\text{B}, 3} - \hat{\Phi}_{\text{B}, 2}
	\label{eq:Results:BackgroundRateEstimation}
\end{equation}

The expected number of background events $N_\text{B}$ can then be calculated as
\begin{equation}
	N_\text{B} = \hat{\Phi}_\text{B}\cdot\Delta t_\text{S}\cdot t_\text{acq}
	\label{eq:Results:BackgroundEventsEstimation}
\end{equation}
where $\Delta t_\text{S}$ is the signal region width from Fig.~\ref{fig:Results:DataSample} and $t_\text{acq}$ is the acquisition time from Table~\ref{tab:Results:DataSets}. Table~\ref{tab:Results:BackgroundSummary} shows the resulting expected number of background events and the observed events for all runs, which were found to be compatible within one standard deviation. It is thus concluded that no excess of zero-energy compatible events was observed at any positron implantation energy.

\begin{table}
	\centering
	\sisetup{separate-uncertainty=false}
	\caption{\label{tab:Results:BackgroundSummary}Expected background events $N_\text{B}$, observed events $N_\text{S}$ and total events $N_\text{tot}$ in the signal region for each implantation energy $E_{\ops}$.}
	\begin{tabular*}{\columnwidth}{c@{\extracolsep{\fill}} S[table-format=4.0(3)] S[table-format=4.0(2)] S[table-format=1.3(1)E1]}
	\toprule   \vspace{-6pt}\\
	$E_{\ops}$ $\left[\si{\kilo\electronvolt}\right]$ & {$N_\text{B}$} & {$N_\text{S}$} & {$N_\text{tot}$} \\
	\hline
	2.0	& 158(36)	& 151(12)	& 2.038(5)E5	\\
	3.0	& 357(55)	& 395(20)	& 3.256(6)E5	\\
	3.0	& 630(130)	& 627(25)	& 6.900(8)E5	\\
	3.5	& 306(32)	& 316(18)	& 1.566(4)E5	\\
	4.0	& 1616(81)	& 1534(39)	& 3.777(6)E5	\\
	\botrule
	\end{tabular*}
	\sisetup{separate-uncertainty=true}
\end{table}

\subsection{\label{subsec:Results:LimitsBr}Limits on Branching Ratios}

Since no signal events were observed above the expected background, upper limits on the branching ratio of the processes $\si{\positron}\to\text{invisible}$ and $\ops\to\text{invisible}$ can be set. In the presence of a known background $N_\text{B}$ and a certain signal $N_\text{S}$, the number of expected events $N_\text{E}$ is
\begin{equation}
	N_\text{E} = N_\text{S} + N_\text{B} = N_\text{tot}\cdot\eta_\text{S}\cdot\mathit{Br} + N_\text{B}
	\label{eq:Results:LimitsBr:AnalysisDefinition}
\end{equation}
where $\eta_\text{S} = \SI{92.1}{\percent}$ is the signal detection efficiency, $N_\text{tot}$ is the total number of events and $\mathit{Br}$ the branching ratio of the process.

The number of observed events can be assumed to follow a Poisson distribution due to the counting nature of the experiment, and all uncertainties are taken to be normally distributed. Using a Bayesian approach with a flat prior distribution, upper limits can be extracted for single and multiple data sets~\cite{Hebbeker2001, VigoPhD}. The resulting limits on the branching ratios are shown in Table~\ref{tab:Results:Results}.

\subsection{\label{subsec:Results:LimitsEpsilon}Limit on Mixing Strength \texorpdfstring{$\varepsilon$}{epsilon}}
Limits on the coupling constant $\varepsilon$ can be extracted from $\text{Br}\left(\ops\to\text{invisible}\right)$. For each implantation energy $E_{\si{\positron}}$, one can assume the corresponding \ops\ mean emission energy extracted from the TOF data~\cite{Crivelli2010_2} and obtain the discrete frequency distribution of collision rate from simulation (Fig.~\ref{fig:ExpSetup:oPsCollisions}). The data for the relevant \ops\ energies are summarized in Table~\ref{tab:ExpSetup:OPsProperties}. Iteratively solving Eq.~\eqref{eq:WeightedAverageBrRatio} yields then the upper limits on $\varepsilon$ shown in Table~\ref{tab:Results:Results}.

\begin{table}
	\sisetup{separate-uncertainty=false}
	\centering
	\caption{\label{tab:Results:Results}Summary of limits on branching ratios $\text{Br}\left(\si{\positron}\to\text{inv.}\right)$  and $\text{Br}\left(\ops\to\text{inv.}\right)$, and resulting limits on the coupling constant $\varepsilon$. All limits are given with \SI{90}{\percent} C.~L.}
	\begin{tabular*}{\columnwidth}{c@{\extracolsep{\fill}} S[table-format=1.2(0)] S[table-format=2.2(0)] S[table-format=1.1(0)]}
		\toprule   \vspace{-6pt}\\
		{\multirow{2}{*}{$E_{\si{\positron}}$ $\left[\si{\kilo\electronvolt}\right]$}} & {$\text{Br}\left(\si{\positron}\to\text{inv.}\right)$} & {$\text{Br}\left(\ops\to\text{inv.}\right)$} & {$\varepsilon$} \\[3pt]
 & {$\left[\num{E-4}\right]$} & {$\left[\num{E-4}\right]$} & {$\left[\num{E-7}\right]$} \\[2pt]
		\hline
		2.0 & 3.2 & 11.2 & 4.6 \\
		3.0 & 4.2 & 15.5 & 5.0 \\
		3.0 & 3.5 & 12.9 & 4.6 \\
		3.5 & 4.8 & 17.8 & 5.2 \\
		4.0 & 3.2 & 12.0 & 4.2 \\
		\hline
		{Combined} & 1.7 & 5.9 & 3.1 \\
		\botrule
	\end{tabular*}
	\sisetup{separate-uncertainty=true}
\end{table}


\section{\label{sec:Conclusions}Conclusions}
In this paper the results of the first search for an invisible decay of \ops\ confined in a vacuum cavity were presented. No event above the expected background was found in the signal region, and thus an upper limit for the branching ratio was obtained:
\begin{equation*}
	\text{Br}\left(\ops\to\text{invisible}\right) < \num{5.9E-4}\quad\left(\SI{90}{\percent}\text{ C.~L.}\right)
\end{equation*}

This result, analyzed in the context of the Mirror Matter model, provides an upper limit on the mirror and ordinary photons kinetic mixing strength
\begin{equation*}
	\varepsilon < \num{3.1E-7}\quad\left(\SI{90}{\percent}\text{ C.~L.}\right)
\end{equation*}
obtained for the first time free of systematic effects due to the absence of collisions of \ops\ with matter.

The main limitation of the experimental sensitivity is the background arising from positron tagging accidentals, which could be overcome with e.g.~a higher positron flux or an improved confinement cavity and tagging system~\cite{Crivelli2010}. Such upgrades would improve the sensitivity to $\varepsilon\sim\num{E-9}$, below the current limit from Big Bang Nucleosynthesis constraints ($\varepsilon\leq\num{3E-8}$), which would confront directly the interpretation of the DAMA/LIBRA claim of a signal observation in terms of Mirror Matter.

\section*{\label{sec:Acknowledgements}Acknowledgements}
We wish to thank S.~Gninenko and A.~Belov for their essential contributions to the first stages of the experiment. We are grateful to A.~Battaglioni and H.~Yarar for their help in the construction of the experimental setup and in the data taking.
We thank R.~Vallery and D.~Cooke for their valuable discussions and helpful comments. This work was supported by the ETH Zurich Grant ETH-35-14-2.

\newpage 
\bibliography{bibliography}

\end{document}